# Viscoelastic Profiling of Rare Pediatric Extracranial Tumors using Multifrequency MR Elastography: A Pilot Study.

C. Metz[1], S. Veldhoen[1], H. E. Deubzer[2], F. Mollica[1], T. Meyer[3], K. Hauptmann[4], A.H. Hagemann[2], A. Eggert[2], I. Sack[3], M.S. Anders[1]

[1]Charité – Universitätsmedizin Berlin, corporate member of Freie Universität Berlin and Humboldt-Universität zu Berlin, Pediatric Radiology, Berlin, Germany

[2]Charité – Universitätsmedizin Berlin, corporate member of Freie Universität Berlin and Humboldt-Universität zu Berlin, Pediatric Hematology and Oncology, Berlin, Germany

[3]Charité – Universitätsmedizin Berlin, corporate member of Freie Universität Berlin and Humboldt-Universität zu Berlin, Radiology, Berlin, Germany

[4]Charité – Universitätsmedizin Berlin, corporate member of Freie Universität Berlin and Humboldt-Universität zu Berlin, Pathology, Berlin, Germany

Corresponding author: Corona Metz, MD (corona.metz@charite.de)

**Abstract**

Objectives: Magnetic resonance elastography (MRE) is a noninvasive technique for assessing the viscoelastic properties of soft biological tissues in vivo, with potential relevance for pediatric tumor evaluation. This study aimed to evaluate the feasibility of multifrequency MRE in children with solid tumors and to report initial findings on stiffness and fluidity across rare pediatric tumor entities. Additionally, the potential of viscoelastic properties as biomarkers of tumor malignancy was explored.

Materials and Methods: Ten pediatric patients (mean age, 5.7 ± 4.8 years; four female) with extracranial solid tumors underwent multifrequency MRE. Shear waves at 30-70 Hz were subsequently generated and measured with a phase-sensitive single-shot spin-echo planar imaging sequence. The obtained shear wave fields were processed by wavenumber (k-)based multi-frequency inversion to reconstruct tumor stiffness and fluidity. The viscoelastic properties within the tumors were quantified and correlated with the apparent diffusion coefficient (ADC). In addition, differences in stiffness and fluidity were assessed across the histopathologically confirmed tumor entities, which were stratified into malignancy-based groups.

Results: MRE was successfully performed in all patients in under five minutes. Differences in viscoelastic properties were observed among tumor entities: Stiffness, fluidity, and their spatial variability increased significantly with tumor malignancy. Furthermore, a significant inverse correlation was observed between stiffness and tumor ADC values.

Conclusion: Multifrequency MRE was feasible in pediatric MRI and provided insight into tumor biomechanics. Preliminary data revealed differences in stiffness and fluidity across pediatric solid tumors correlating with malignancy. MRE holds promise for diagnosis and classification of pediatric tumor entities and their malignancy.

**Abbreviations**

| | |
|---|---|
| ADC | apparent diffusion coefficient |
| D-SNR | displacement signal-to-noise ratio |
| DWI | diffusion-weighted imaging |
| FOV | field of view |
| GRAPPA | generalized autocalibrating partially parallel acquisitions |
| MEG | motion encoding gradient |
| MRE | magnetic resonance elastography |
| MRI | magnetic resonance imaging |
| SWS | shear wave speed |
| TE | time of echo |
| TR | time of repetition |

**Introduction**

Magnetic resonance elastography (MRE) is a non-invasive imaging technique used to quantify viscoelastic properties of biological soft tissues in vivo (1). Mechanical waves are generated, and motion-sensitive magnetic resonance imaging (MRI) sequences monitor the propagation of the resulting shear waves (2). The measured shear wave characteristics are typically processed by wave inversion algorithms to generate viscoelastic parameter maps (3). This technique provides high spatial resolution enabling detailed assessment of biomechanical properties without the need for invasive procedures, for example, in the evaluation of liver fibrosis or cirrhosis (4-6).

Enzymes, such as lysyl oxidases, influence the formation of collagen fibers and their cross-linking. Increased cross-linking and thus stiffness of the tissue is observed in various fibrotic tissues and in desmoplastic reactions within tumors and is associated with aggressive cancer phenotypes (7, 8). Furthermore, it has been speculated that cancer cell motility influences the macroscopic tissue fluidity as measured by MRE in tumors (9). Therefore, MRE could add diagnostic information to pediatric tumor differentiation, therapy response and outcome by determining the viscoelasticity of tumor tissue (10).

Diffusion-weighted MRI (DWI) is routinely used in clinical routine to distinguish between benign and malignant lesions by assessing the diffusion behavior of water molecules within tissue (11). By employing diffusion-sensitizing gradients, DWI captures microscopic water displacement over distances of approximately 1–20 µm. In oncologic imaging, restricted diffusion, typically due to increased cellular density and reduced extracellular space, is a hallmark of malignant tumors (12-14). Given these characteristics, correlating DWI with MRE-derived parameters may provide complementary insights in tumor grading.

The diagnostic value of stiffness and fluidity as biomarkers in oncological imaging has been demonstrated in adults through pilot studies (15, 16). However, in the pediatric population, the potential of MRE remains largely unexplored. To date, only a few studies have investigated MRE in children, primarily focusing on the liver and the brain, with just one oncological study in children on intracerebral gliomas (17-20). No data are available on MRE of extracranial solid tumors in children, leaving its diagnostic potential in this context undefined. This pilot study aimed to assess the feasibility of multifrequency MRE in rare pediatric extracranial tumors and to explore its diagnostic potential. Therefore, we present initial results of MRE in different pediatric tumor entities.

## Materials and Methods

*Institutional Review*

The prospective study was approved by the institutional review board (*Charité – Universitätsmedizin Berlin*). Written informed consent of the legal guardian was obtained prior to the examinations.

*Study sample and design*

Inclusion criteria were newly or recently diagnosed extracranial pediatric solid tumors. Depending on the patient's age and compliance, the examinations were performed under anesthesia (n = 5) or with the patient awake (n = 5). Exclusion criteria were refusal by the legal guardians (n = 0) and premature cancellation of the examination due to insufficient sedation (n = 1).

*Study population*

Ten pediatric patients (mean age, 5.7 ± 4.8 years; median age, 4.7 years; age range, 4 month – 15 years; four female) with different tumor entities were included into this prospective single-center study:

- one patient with abdominal alveolar rhabdomyosarcoma
- one patient with Ewing's sarcoma arising from the first rib
- one patient with telangiectatic osteosarcoma of the left proximal tibial metaphysis
- one patient with hepatoblastoma in the left liver lobe
- one patient with nephroblastoma of the left kidney
- three patients with neuroblastomas:
    - one with high-risk neuroblastoma of the right adrenal gland
    - one with cystic low-risk neonatal neuroblastoma (stage MS) in both adrenal glands
    - one with low-risk neuroblastoma of the left adrenal gland
- one patient with schwannoma on the proximal right forearm
- one patient with lipoma on the right elbow

All patients were examined using a 3-Tesla MRI scanner (Magnetom Skyra, Siemens Healthineers, Germany) using a 12-channel receiver coil.

Tumors were risk-categorized based on biological factors, including tumor grading, MYCN amplification, proliferation rate, and gene expression profiles, as well as clinical features such as presence of metastases, prognosis, expected therapy response, and recurrence rate (21-27). Tumors were manually segmented by a trained radiologist (C.M.) with six years of experience in radiology on gradient echo morphologic images. The segmentation was resliced to match the MRE and DWI voxel grids using elastix (28) with an identity transform and nearest-neighbor interpolation.

*MRE*

Vibrations were generated by four custom-made compressed air drivers attached to the patient using a Velcro belt positioned near the tumor localization. The compressed air flow was modulated to induce rhythmic in- and deflation of the drivers at the target frequency, resulting in time harmonic tissue displacements. Data acquisition was synchronized with vibrations by calibrating the internal clocks of the MRI system and the vibration generator. The pediatric MRE setup is shown in *Figure 1*.

Vibrations were consecutively generated at 30, 40, 50, 60, and 70 Hz. Each corresponding wave period was sampled at eight equidistant temporal phases using a nullified first-order motion encoding gradient (MEG). The amplitude of the MEG was set to 40 mT/m with a duration of 15.20 ms, resulting in encoding efficiencies of 27.4, 17.1, 12.6, 10.3, and 9.4 rad/mm for the vibration frequencies of 30 to 70 Hz, respectively. MEGs were played out along the three Cartesian axes of the imaging coordinate system.

A total of 24 slices were acquired in a transverse orientation using a single-shot spin-echo echo-planar imaging sequence. Field of view (FOV) options of 180×140 or 280×224 mm² were selected based on patient size, with acquisition times of 4:20 or 4:31 min, respectively. Imaging parameters were defined as follows: Voxel size = 2.0×2.0×5.0 mm³; time of repetition (TR) = 2100 or 2150 ms; time of echo (TE) = 44 ms; GRAPPA = 2; phase partial Fourier = 7/8.

The obtained MRE data were processed using the wavenumber (k-)based multi-frequency dual elasto visco inversion (k-MDEV), available as an online tool (3), to reconstruct stiffness maps in terms of shear wave speed (SWS in m/s) and phase angle of the complex shear modulus (Phi in rad), also termed tissue fluidity.

*Quality Control for Viscoelastic Parameter Estimation*

Measurement success was assessed using the displacement signal-to-noise ratio (D-SNR), averaged across all applied excitation frequencies. Measurements with a mean D-SNR above 38.18 were considered acceptable for further analysis. For statistical analysis, only regions with displacement amplitudes greater than 3.6 µm were included, as they indicate accurate estimation of the underlying viscoelastic properties (29).

*DWI*

DWI was performed using a gradient-echo echo-planar imaging sequence. Data were acquired at b-values of 50 and 800 s/mm² with 4 and 16 repetitions, respectively, using 12 diffusion-encoding directions. Depending on the size of the patient and the body parts examined, the FOV ranged from 168×280 to 310×382 mm² with 42-120 slices. Imaging parameters were defined as follows: Voxel size = 1.5×1.5×4.0 mm³; TR = 2000 ms; TE = 55-108 ms; Simultaneous Multi Slice Factor = 2.

*Statistics*

Statistical analyses were performed using Matlab (version R2024a, MathWorks Inc, United States). The viscoelastic properties of the tumors were correlated with their ADC and ordinal risk groups using Pearson correlation and Spearman's rank correlation, respectively. P-values < 0.05 were considered statistically significant.

**Results**

Multifrequency MRE was performed successfully and without complications in all patients.

Table 1 shows the study population, and tumor entities with associated stiffness, fluidity, ADC and measurement quality metrics. Tumors were located in various anatomical regions, including the abdomen, thorax, and extremities. All measurements yielded displacement signal-to-noise ratios within the tumor regions exceeding the threshold of 38.18 for mean D-SNR and were therefore classified as successful acquisitions. In all cases except for the Ewing´s sarcoma and the rhabdomyosarcoma, at least 99% of voxels within the demarcated tumor regions exceeded the displacement amplitude threshold of 3.6 µm. The rhabdomyosarcoma exhibited over 96% of voxels above this threshold, while the Ewing sarcoma showed a markedly lower proportion, with only 54% of voxels meeting the criterion.

Figure 2 shows representative slices of MRE magnitude, SWS, fluidity, and ADC for each tumor entity investigated. Yellow contours denote the tumor boundaries where sufficient shear wave displacement was present. The displacement fields show adequate wave propagation coverage within these regions, supporting the reliability of subsequent mechanical parameter estimation. Across tumor entities, a general trend is observed: more biological and clinical aggressive entities, such as rhabdomyosarcoma and high-risk neuroblastoma, exhibit higher SWS and lower ADC values, consistent with increased tissue stiffness and restricted diffusivity. In contrast, benign or low-malignancy tumors such as lipoma and low-risk neonatal neuroblastoma show lower SWS and higher ADC values.

Figure 3 presents SWS, fluidity, and ADC with their standard deviations, alongside tumor risk groups. Figures 3A and 3B display SWS and fluidity across risk groups, respectively, while Figure 3C shows ADC as a function of SWS. This visualization highlights distinct mechanical profiles among tumor types, with higher SWS, fluidity, and corresponding intratumoral variability observed in high-risk tumors.

Stiffness and fluidity showed a significant positive correlation with tumor risk group ($\rho$ = 0.72, $p < 0.05$) and ($\rho$ = 0.84, $p < 0.01$), respectively. When the telangiectatic osteosarcoma was excluded (significant differences in the viscoelastic properties of the tumor due to the large proportion of partially hemorrhaged cysts), stiffness and fluidity remained positively correlated with tumor risk group, whereas the correlation was higher for stiffness ($\rho = 0.93$, $p < 0.01$) and lower for fluidity ($\rho = 0.80$, $p < 0.05$).

Intratumoral viscoelastic heterogeneity also correlated significantly positively with tumor risk group, with stiffness heterogeneity ($\rho = 0.85$, $p < 0.01$) and fluidity heterogeneity ($\rho = 0.88$, $p < 0.001$). When the telangiectatic osteosarcoma was excluded, the correlation increased to ($\rho = 0.88$, $p < 0.01$) for stiffness heterogeneity and was ($\rho = 0.87$, $p < 0.01$) for fluidity heterogeneity. These results further support the association between viscoelastic properties and tumor aggressiveness.

A significant negative correlation was observed between SWS and ADC ($r = -0.71$, $p < 0.05$), when the lipoma and hepatoblastoma was excluded (diffusion-weighted imaging is unreliable in fatty lesions and the treated state of the hepatoblastoma requires separate consideration, see discussion), while no significant correlation was found between fluidity and ADC.

**Discussion**

To the best of our knowledge, this is the first study to investigate viscoelastic tissue characteristics of pediatric extracranial solid tumors using MRE, highlighting its potential as a non-invasive imaging biomarker in pediatric oncology.

All pediatric MRE scans surpassed the necessary displacement signal-to-noise ratio threshold of 38.18 within the tumor regions to ensure successful measurements. Most tumors showed near-complete voxel coverage above the displacement amplitude criterion of 3.6 µm, delineating robust shear wave propagation areas. The notably lower proportion of reliable voxels in the Ewing´s sarcoma is likely due to its deep anatomical location, where shear wave penetration is attenuated, leading to areas with reduced displacement amplitudes. This agrees with a known limitation of externally induced wave propagation in deeper or acoustically shielded regions (30), though it did not affect overall diagnostic feasibility. The MRE quality metrics verified successful integration of the technique into the diagnostic workflow without complications for all pediatric patients, with acquisition times consistently under five minutes and no reported complications. Our findings are in good agreement with those of a large study by Joshi et al. (31), in which 96% of MRE examinations in children and young adults were successfully completed.

The preliminary findings of our study on MRE of extracranial solid tumors in children revealed strong correlations between viscoelastic properties and tumor malignancy. Stiffness and fluidity increased with higher tumor risk, with consistently high correlation coefficients, highlighting MRE as a promising tool for non-invasive tumor staging in pediatric tumors. Intratumoral viscoelastic heterogeneity also correlated significantly with tumor risk, further underscoring the sensitivity of MRE to tumor microstructural complexity. Additionally, SWS showed a significant inverse correlation with ADC, consistent with the expected relationship between stiffness and cellular density.

The rhabdomyosarcoma exhibited the highest stiffness and tissue fluidity, while the two benign tumors in our cohort—a schwannoma and a lipoma—showed the lowest values. Ewing's sarcoma being less stiff than the rhabdomyosarcoma might be attributable to its anatomical location in the first left rib, where surrounding compliant thoracic tissues (muscle, lung, pleura) may mechanically buffer the tumor and reduce measurable stiffness. Additionally, Ewing's sarcoma is known to be a highly vascularized tumor (32, 33), which may also contribute to lower stiffness and increased fluidity. SWS values of the hepatoblastoma being similar to the biologically very aggressive tumor types, may be artificially elevated due to chemotherapy-induced tumor regression and associated fibrotic remodeling. The telangiectatic osteosarcoma exhibited low stiffness and fluidity values compared to the other sarcomas and highly malignant tumors in our cohort. This may be explained by reduced extracellular matrix content, low cellular density, or limited stromal organization, all of which contribute to decreased mechanical integrity (34). As a result, the measured viscoelastic properties may not accurately reflect the biological aggressiveness of this tumor entity but rather the physical properties of its cystic architecture. This is supported by stiffness correlating more closely with tumor risk group than fluidity when the telangiectatic osteosarcoma is excluded.

Cystic neuroblastoma and nephroblastoma, both of which contain substantial fluid-filled components, demonstrated stiffness and fluidity values higher than those of benign tumors but lower than those of other malignant lesions. High-risk neuroblastoma exhibited higher viscoelastic properties compared to the two low-risk neuroblastomas, reflecting its more aggressive biological behavior. Low-risk neuroblastomas typically have an excellent prognosis and can, in some cases, such as stage MS, be managed with observation due to their potential for spontaneous regression (35). In contrast, high-risk neuroblastomas are associated with poor outcomes and require multimodal therapy (36, 37). The two low-risk neuroblastomas differed only slightly, with the cystic lesion showing lower stiffness and higher fluidity, likely attributable to the physical properties of cystic fluids. These findings suggest that viscoelastic properties may serve as potential imaging biomarkers for risk stratification in pediatric neuroblastoma.

The significant inverse correlation between DWI-derived ADC values and MRE parameters supports the notion that MRE may be a valuable tool for assessing tumor malignancy, as both techniques reflect underlying tissue architecture: increased cellularity and reduced extracellular space in malignant tumors lead to lower ADC values and obviously higher tissue stiffness.

The ADC map of the lipoma was excluded from analyses due to the unreliability of DWI in fat-rich tissues. This limitation arises from the low water content and distinct diffusion properties of fat, which result in weak diffusion signals and poor contrast. Additionally, fat suppression techniques—commonly employed to enhance DWI quality—may be insufficient in these tissues (38).

In future studies, it may also be valuable to explore correlations between MRE-derived parameters and additional imaging biomarkers, such as those from dynamic contrast-enhanced (DCE) MRI, to gain further insights into tumor vascularity, perfusion dynamics, and their relationship to tissue biomechanics.

*Study limitations*

The current findings represent preliminary data and individual observations. Further studies involving larger pediatric cohorts of certain tumor entities are needed to validate these results and expand on the utility of MRE in tumor characterization, treatment monitoring, and prognostication.

*Conclusion*

Multifrequency MRE was readily integrated into the routine diagnostic workflow of pediatric MRI and provided valuable insights into the mechanical properties of rare tumor entities. Our preliminary results reveal that stiffness and fluidity vary across different pediatric tumor types and correlate with malignancy, indicating that MRE is a promising technique for the diagnosis and classification of pediatric solid tumors. Moreover, serial MRE imaging may help monitor treatment-induced changes in tissue biomechanics, and MRE-derived parameters could contribute to future approaches for therapy response assessment. However, in tumors with large cystic or necrotic components, MRE measurements may be limited or biased and should therefore be interpreted with caution, at least until larger studies can more clearly define their diagnostic reliability in such contexts.

**Tables and table legends**

Table 1: Study population, tumor characteristics, and obtained imaging parameters. ADC: apparent diffusion coefficient; SWS: shear wave speed; Phi: fluidity; D-SNR: displacement signal-to-noise ratio; DA: displacement amplitude. Imaging metrics are presented as mean ± standard deviation. No ADC value is provided for the lipoma, as diffusion-weighted imaging is unreliable in fatty lesions due to the low free water content and potential confounding effects from inadequate fat suppression.

| Tumor type | Location | Risk Group | Age in years / months | Sex | SWS in m/s | Phi in rad | D-SNR | DA in µm | Reliable Voxel in % | ADC in ×10-3mm²/s | Comment |
|---|---|---|---|---|---|---|---|---|---|---|---|
| Rhabdomyosarcoma | Intraabdominal | 4 | 9 / 10 | male | 3.20± 1.16 | 0.84 ± 0.32 | 53.9 ± 0.4 | 6.6 ± 3.8 | 95 | 0.90 ± 0.45 | Metastatic disease, treatment-naive |
| Ewing´s sarcoma | First left rib | 4 | 6 / 8 | female | 2.30 + 0.62 | 1.12 ± 0.40 | 44.3 ± 0.9 | 4.1 ± 2.4 | 54 | 1.02 ± 0.53 | Treatment-naive |
| High-risk neuroblastoma | Right adrenal gland | 4 | 1 / 6 | male | 2.50 ± 1.21 | 1.15 ± 0.42 | 43.4 ± 1.8 | 5.9 ± 2.9 | 100 | 1.12 ± 0.43 | Treatment-naive |
| Telangiectatic osteosarcoma | Left proximal tibial metaphysis | 4 | 15 / 11 | female | 1.75 ± 0.71 | 0.74 ± 0.32 | 38.95 ± 4.9 | 8.5 ± 5.9 | 95 | 1.12 ± 0.63 | Treatment-naive |
| Hepatoblastoma | Liver segments II-IV | 3 | 1 / 9 | male | 2.36 ± 0.71 | 0.60 ± 0.25 | 46.3 ± 2.7 | 16.5 ± 8.1 | 100 | 1.50 ± 0.80 | Neoadjuvant chemotherapy, treatment-naive |
| Nephroblastoma | Left kidney | 3 | 2 / 11 | female | 2.15 ± 1.04 | 0.68 ± 0.28 | 45.2 ± 2.0 | 9.5 ± 6.2 | 99 | 1.38 ± 0.71 | Metastatic disease, treatment-naive |
| Low-risk neuroblastoma | Left adrenal glands | 2 | 1 / 6 | female | 1.92 ± 0.51 | 0.52 ± 0.24 | 54.8 ± 2.6 | 11.5 ± 7.4 | 100 | 1.18 ± 0.45 | Treatment-naive |
| Low-risk neonatal neuroblastoma | Both adrenal gland | 2 | 0 / 4 | male | 1.87 ± 0.59 | 0.68 ± 0.29 | 41.5 ± 1.4 | 11.0 ± 5.9 | 100 | 1.16 ± 0.60 | Metastatic disease, Stadium MS, treatment-naive |

| Schwannoma | Right forearm | 1 | 8 / 0 | male | 1.58 ± 0.31 | 0.54 ± 0.14 | 38.4 ± 3.8 | 21.1 ± 10.5 | 100 | 1.38 ± 0.34 | Treatment-naive |
| Lipoma | Right elbow | 1 | 8 / 1 | male | 1.33 ± 0.27 | 0.66 ± 0.25 | 38.9 ± 1.3 | 10.2 ± 5.5 | 99 | - | Treatment-naive |

**Figures and figure legends**

*Figure 1* Pediatric MRE setup. The pressure control box controles the pressurized air flow used to periodically inflate and deflate four compressed air drivers at the desired frequency. The air drivers are attached near the region of interest using a Velcro belt (In the example shown, posterior and anterior of the abdomen). The resulting periodic tissue displacement alters the phase of the MR signal, which is captured by the radiofrequency receiver coil.

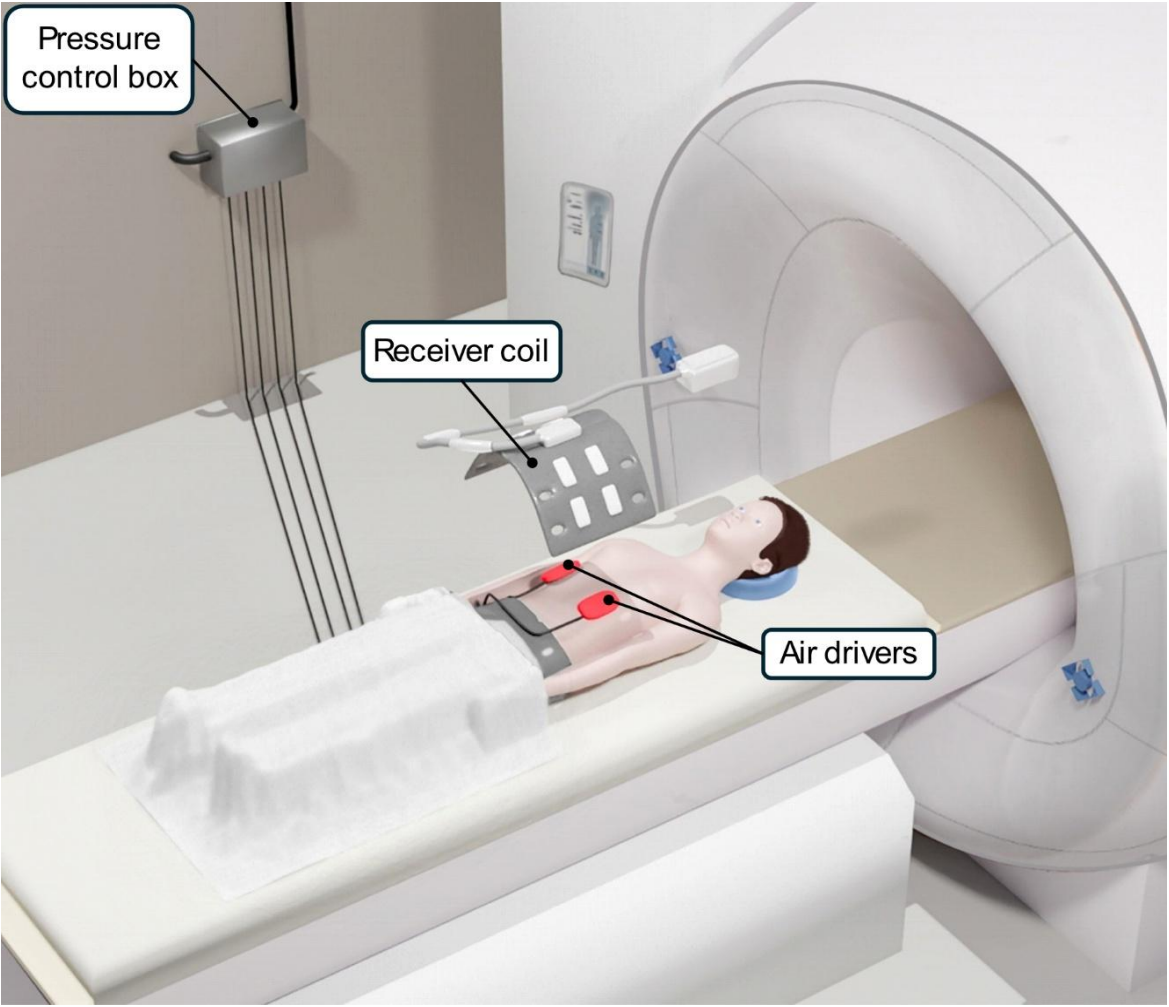

*Figure 2* Representative slices of MRE magnitude, shear wave speed, viscosity-related fluidity, and ADC maps of each tumor. Tumor margins as delineated by a trained radiologist are shown, with regions exceeding the displacement threshold of 3.6 µm demarcated by yellow lines.

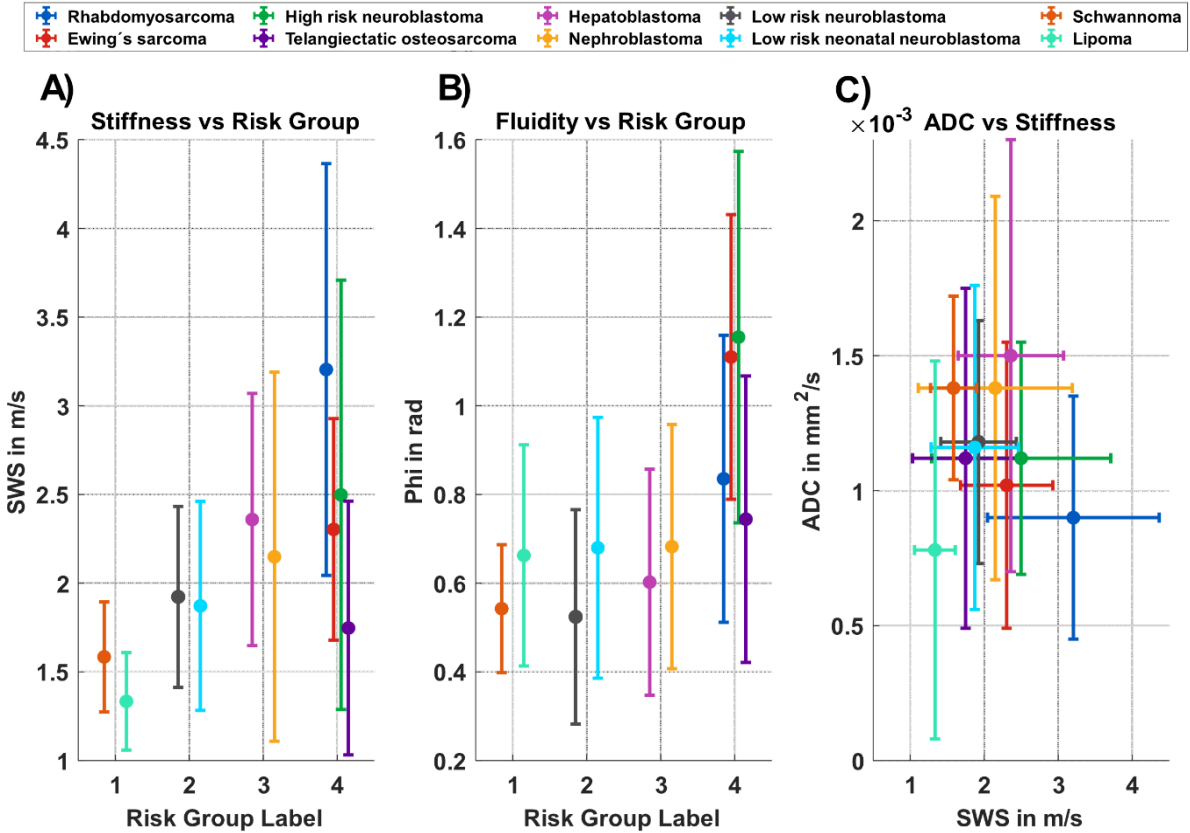

*Figure 3* Viscoelastic and diffusion properties of pediatric tumors. A) and B) illustrate the correlation of stiffness and fluidity, respectively, across clinical and biological risk groups, revealing trends with increasing risk. C) shows the inverse relationship between stiffness and apparent diffusion coefficient (ADC), reflecting the mechanical–diffusional cross-dependence across tumor types. Each color represents a different tumor entity. Horizontal offsets were applied in A) and B) for visual clarity. Error bars indicate the spatial standard deviation within the tumor volume.

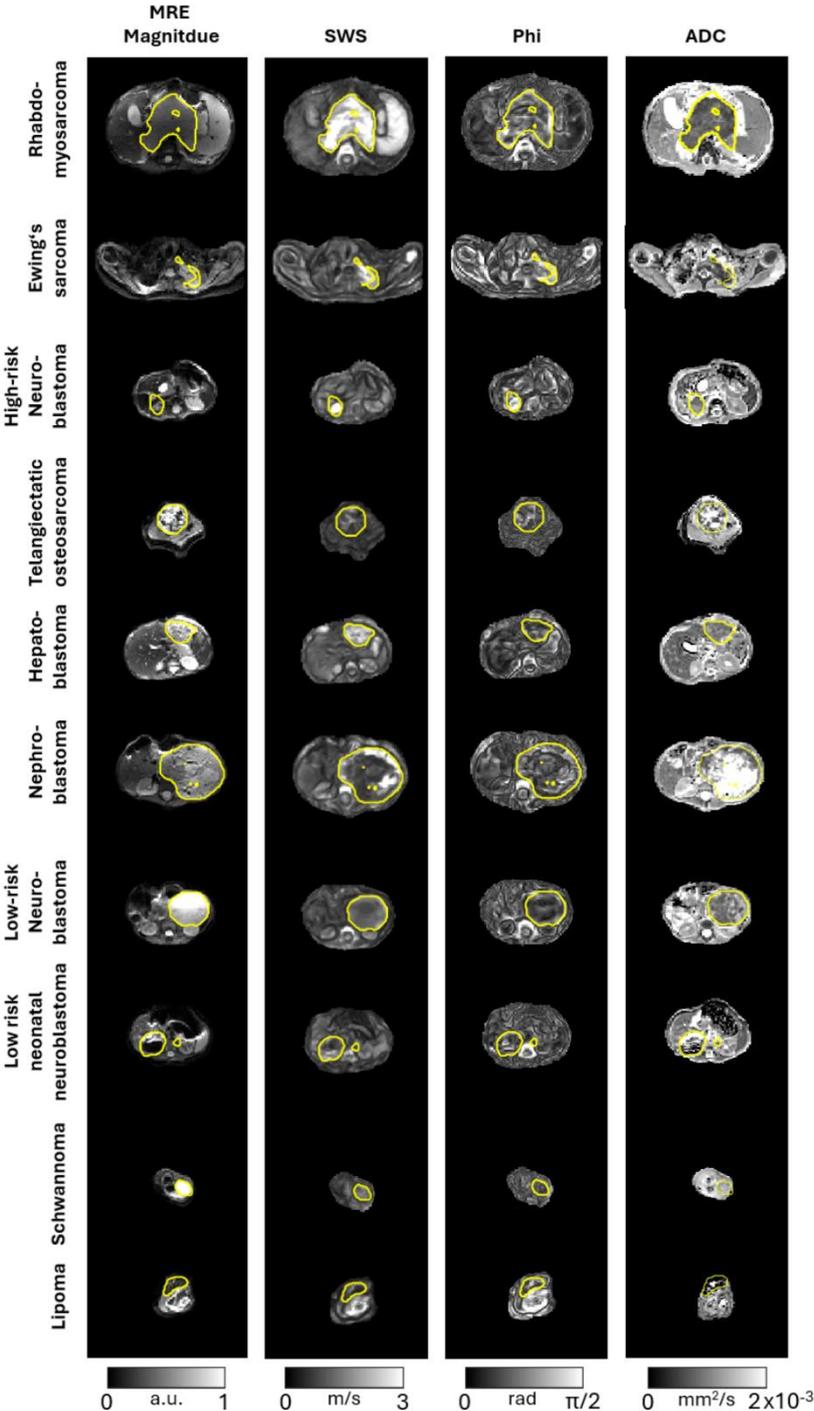